\documentclass[tradiabstract]{aa} 
\usepackage{graphicx}
\usepackage{txfonts}
\usepackage{natbib}
\begin{document}
%
   \title{Photospheric activity, rotation, and radial velocity variations of the planet-hosting star CoRoT-7\thanks{Based on observations obtained with CoRoT, a space project operated by the French Space Agency, CNES, with partecipation of the Science Programme of ESA, ESTEC/RSSD, Austria, Belgium, Brazil, Germany, and Spain.}} 
  
\authorrunning{A. F. Lanza et al.}
\titlerunning{Activity, rotation, and radial velocity variations of CoRoT-7}

   \author{A.~F.~Lanza\inst{1} \and A.~S.~Bonomo\inst{1,2} \and C.~Moutou\inst{2}
          \and I.~Pagano\inst{1} \and S.~Messina\inst{1} \and G.~Leto\inst{1} \and G.~Cutispoto\inst{1}
          \and S.~Aigrain\inst{3} \and R.~Alonso\inst{4} \and P.~Barge\inst{2}, M.~Deleuil\inst{2}  
         \and M.~Auvergne\inst{5} \and A.~Baglin\inst{5} 
       \and A.~Collier~Cameron\inst{6}  
          }

   \institute{INAF-Osservatorio Astrofisico di Catania, Via S.~Sofia, 78, 95123 Catania, Italy\\
              \email{nuccio.lanza@oact.inaf.it}
\and 
Laboratoire d'Astrophysique de Marseille (UMR 6110),
Technopole de Ch\^{a}teau-Gombert,
38 rue Fr\'ed\'eric Joliot-Curie,
13388 Marseille cedex 13, France     
\and Department of Physics, University of Oxford, Denys Wilkinson Building, Keble Road, Oxford OX1 3RH, United Kingdom
\and
Observatoire de Gen\`eve, Universit\'e de Gen\`eve, 51 Ch. des Maillettes, 1290, Sauverny, Switzerland  
\and 
    LESIA, CNRS UMR 8109, Observatoire de Paris, 5 place J. Janssen, 92195 Meudon, France       
    \and 
    School of Physics and Astronomy, University of St. Andrews, 
    North Haugh, St Andrews, Fife  KY16 9SS, Scotland   
    }

   \date{Received ; accepted }

 
  \abstract
   {The CoRoT satellite has recently discovered the transits of a telluric planet across the disc of a late-type magnetically active star dubbed CoRoT-7, while a second planet has been detected {{ after filtering out the radial velocity (hereafter RV) variations due to stellar activity.}} }
   { {{ We investigate the magnetic activity of CoRoT-7 and use the results for a better understanding of the impact of magnetic activity on stellar RV variations. }} }
   {{{ We derive the longitudinal distribution of active regions on CoRoT-7 from a maximum entropy spot model of the CoRoT light curve. Assuming that each active region consists of dark spots and bright faculae in a fixed  proportion, we synthesize the expected RV variations. }} }
   {Active regions are mainly located at three active longitudes which {{ appear to }} migrate at different rates, probably as a consequence of surface differential rotation, {{ for which a lower limit of $\Delta \Omega / \Omega = 0.058 \pm 0.017$ is found}}. The synthesized activity-induced RV variations  reproduce the  amplitude of the observed RV curve and are used to study the impact of stellar activity on planetary detection.  } 
  {{ {{In spite of the non-simultaneous CoRoT and HARPS observations, our study confirms 
 the validity of the method previously adopted to filter out RV variations induced by stellar activity. We find a false-alarm probability $ < 10^{-4}$ that the RV oscillations attributed to CoRoT-7b and CoRoT-7c are spurious effects of noise and activity. Additionally, our model suggests that other periodicities found in the observed RV curve of CoRoT-7 could be explained by active regions  whose visibility is modulated by a differential stellar rotation with periods ranging from 23.6 to 27.6 days.}}  }}


   \maketitle
%

\section{Introduction}

CoRoT (Convection, Rotation and Transits) is a photometric space experiment devoted to asteroseismology and the search for extrasolar planets by the method of transits \citep{Baglinetal06}. It has recently discovered two telluric planets around a late-type star called CoRoT-7, the innermost of which transits across the disc of the star and has an orbital period of only 0.8536 days, while the other has a period of 3.69 days and does not show transits \citep{Legeretal09,Quelozetal09}. The mass of the inner planet CoRoT-7b is $4.8 \pm 0.8$ M$_{\oplus}$, while  that of CoRoT-7c is $8.4 \pm 0.9$ M$_{\oplus}$, assuming they are on coplanar orbits \citep{Quelozetal09}. The relative depth of the transit of CoRoT-7b is $\Delta F/F = (3.35 \pm 0.12) \times 10^{-4}$ in the CoRoT white passband (see Sect.~\ref{observations}), leading to a planetary radius of $1.68 \pm 0.09$ R$_{\oplus}$. 

CoRoT-7 is an active star showing a rotational modulation of its optical flux with an amplitude up to $\sim 0.03$ mag {and} a period of $\sim 23.5$ days \citep{Quelozetal09}. 
The discovery of CoRoT-7c and the measurement of the mass of CoRoT-7b have become possible with the HARPS  spectrograph after filtering out the apparent variations of the stellar RV induced by magnetic activity which have an amplitude of $\approx 40$ m s$^{-1}$, i.e., $5-6$ times greater than the wobbles produced by the gravitational pull of the planets. 
{Therefore, a more detailed model of
the stellar activity is needed to improve RV measurements
and to obtain better planetary parameters.
This would also rule
out the possibility that a quasi-periodic RV signal is
due to rotating spots on the stellar surface instead of a
planetary companion as, e.g., in the case of HD~166435 \citep{Quelozetal01}. }

We apply the same spot modelling approach used for the high-precision light curves of CoRoT-2 
\citep{Lanzaetal09a} and CoRoT-4 \citep{Lanzaetal09b}, deriving the longitudinal distribution of photospheric brightness inhomogeneities and their evolution {during} the $\sim 130$ days of CoRoT observations. The different rotation rates of active regions  allow us to estimate a lower limit for the amplitude of  the differential rotation of CoRoT-7. From our spot maps we  synthesize the activity-induced RV perturbations using a simple model for the distortion of the line profiles, as described in Sect.~\ref{line_prof_model}. A Fourier analysis of such a synthetic time series  {yields typical periods} induced by stellar activity, allowing us to discuss its impact on the detection of planets around CoRoT-7.

\section{Observations}
\label{observations}

CoRoT-7 has been observed during the first long run of CoRoT toward the galactic anticentre from 24 October 2007 to 3 March 2008. Since the star is bright ($ V =11.67$), the time sampling is 32~s from the beginning of the observations. CoRoT performs aperture photometry with a fixed mask \citep[see Fig.~6 in ][]{Legeretal09}. The contaminating  flux from background stars falling inside the mask amounts to a maximum of 0.9 percent and produces a dilution of the photometric variation of CoRoT-7 lower than $1.8 \times 10^{-4}$ mag that can be safely ignored for our purposes \citep[see ][ for upper limits on the background flux variations]{Legeretal09}. The flux inside the star's mask is split along detector column boundaries into broad-band red, green, and blue channels.

The observations and data processing are described by \citet{Legeretal09}, to whom we refer the reader for details. The reduction pipeline applies corrections for the sky background and the pointing jitter of the satellite, which is particularly relevant during  ingress and  egress from the Earth shadow. Measurements during the crossing of the South Atlantic Anomaly of the Earth's magnetic field, which amounts to about $7-9$ percent of {each} satellite orbit, are discarded. High-frequency spikes due to cosmic ray hits and outliers are removed by applying a 7-point running mean. The final duty cycle of the 32-s observations is 88.8 percent for the so-called N2 data time series that are accessible through the CoRoT Public Data Archive at IAS ({http://idoc-corot.ias.u-psud.fr/}). More information on the instrument, its operation, and performance can be found in \citet{Auvergneetal09}.

 To increase the signal-to-noise ratio and reduce systematic drifts present in individual channels, we sum up the flux in the red, green, and blue channels to obtain a  light curve in a spectral range extending from 300 to 1100 nm. 

The transits are removed using the ephemeris of \citet{Legeretal09} and the out-of-transit data are binned by computing {average flux values} along each orbital period of the satellite (6184~s). This has the advantage of removing tiny systematic variations associated with the orbital motion of the satellite \citep[cf. ][]{Alonsoetal08,Auvergneetal09}.  In such a way we obtain a  light curve consisting of 1793 {average points}  ranging from HJD 2454398.0719 to HJD 2454528.8877, i.e., with a duration of 130.8152 days. The average standard deviation of the points is $2.678 \times 10^{-4}$ in relative flux units. The  maximum observed flux in the average point time series at HJD~2454428.4232 is adopted as a reference unit level corresponding to the unspotted star, since the true value of the unspotted flux is unknown.

\section{Light curve and radial velocity modelling}

\subsection{Spot modelling of wide-band light curves}

\label{spotmodel}

The reconstruction of the surface brightness distribution from the rotational modulation of the stellar flux is an ill-posed problem, because the variation of the flux vs. rotational phase contains {only} information  on the distribution of the brightness inhomogeneities vs. longitude. The integration over the stellar disc effectively cancels any latitudinal information, particularly when the inclination of the rotation axis along the line of sight is close to $90^{\circ}$, as it is assumed in the present case \citep[see Sect.~\ref{model_param} and ][]{Lanzaetal09a}. Therefore, we need to include a priori information in the light curve inversion process to obtain a unique and stable map. This is  done by computing a maximum entropy ({hereafter} ME) map, which has been proven {to successfully} reproduce active region distribution and area variations in the case of the Sun \citep[cf. ][]{Lanzaetal07}. 

In our model, the star is subdivided into {200 surface elements, namely  200  squares} of side $18^{\circ}$, with  each element containing unperturbed photosphere, dark spots, and facular areas. The fraction of an element covered by dark spots is indicated by the filling factor $f$,  the fractional  area of the faculae is $Qf$, and the fractional area of the unperturbed photosphere is $1-(Q+1)f$. 
The contribution to the stellar flux coming from the $k$-th surface element at the time $t_{j}$, where $j=1,..., N$,  is an index numbering the $N$ points along the light curve, is given by:
\begin{eqnarray}
\Delta F_{kj} & = & I_{0}(\mu_{kj}) \left\{ 1-(Q+1)f + c_{\rm s} f +  \right. \nonumber \\
  & & \left.  Q f [1+c_{\rm f} (1 -\mu_{kj})] \right\} A_{k} \mu_{kj} {w}(\mu_{kj}),
\label{delta_flux}
\end{eqnarray}
where $I_{0}$ is the specific intensity in the continuum of the unperturbed photosphere at the isophotal wavelength of the observations, $c_{\rm s}$ and $c_{\rm f}$ are the spot and facular contrasts, respectively \citep[cf. ][]{Lanzaetal04}, $A_{k}$ is the area of the $k$-th surface element,
\begin{equation}
 {w} (\mu_{kj}) = \left\{ \begin{array}{ll} 
                      1  & \mbox{if $\mu_{kj} \geq 0$}  \\
                      0 & \mbox{if $\mu_{kj} < 0$ }
                              \end{array} \right. 
\end{equation}
is its visibility, and 
\begin{equation}
\mu_{kj} \equiv \cos \psi_{kj} = \sin i \sin \theta_{k} \cos [\ell_{k} + \Omega (t_{j}-t_{0})] + \cos i \cos \theta_{k},
\label{mu}
\end{equation}
is the {cosine} of the angle $\psi_{kj}$ between the normal to the surface element and the direction of the observer, with $i$ being the inclination of the stellar rotation axis along the line of sight, $\theta_{k}$ the colatitude and $\ell_{k}$ the longitude of the $k$-th surface element; $\Omega$ {denotes} the angular velocity of rotation of the star ($\Omega \equiv 2 \pi / P_{\rm rot}$, with $P_{\rm rot}$ the stellar rotation period), and $t_{0}$ the initial time. The specific intensity in the continuum varies according to a quadratic limb-darkening law, as adopted by \citet{Lanzaetal03} for the case of the Sun, viz. $I_{0} \propto a_{\rm p} + b_{\rm p} \mu + c_{\rm p} \mu^{2}$. The stellar flux computed at the time $t_{j}$ is then: $F(t_{j}) = \sum_{k} \Delta F_{kj}$. To warrant a relative precision of the order of $10^{-5}$ in the computation of the flux $F$, each surface element is further subdivided into {$1^{\circ} \times 1^{\circ}$-elements } and their contributions, calculated according to Eq.~(\ref{delta_flux}), are summed up at each given time to compute the contribution of the $18^{\circ} \times 18^{\circ}$ surface element to which they belong.  

We fit the light curve by {varying} the value of $f$ over the surface of the star, while $Q$ is held constant. Even fixing the rotation period, the inclination, and the spot and facular contrasts \citep[see ][ for details]{Lanzaetal07}, the model has 200 free parameters and suffers from  non-uniqueness and instability. To find a unique and stable spot map, we apply ME regularization, as described in \citet{Lanzaetal07}, by minimizing a functional $Z$, which is a linear combination of the $\chi^{2}$ and  the entropy functional $S$; i.e.,
\begin{equation}
Z = \chi^{2} ({\vec f}) - \eta S ({\vec f}),
\end{equation}
where ${\vec f}$ is the vector of the filling factors of the surface elements, $\eta > 0$   a Lagrangian multiplier determining the trade-off between light curve fitting and regularization; the expression for $S$ is given in \citet{Lanzaetal98}.  The entropy functional $S$ is constructed in such a way that it attains its maximum value when the star is {free of spots}. Therefore, by increasing the Lagrangian multiplier $\eta$, we increase the weight of $S$ in the model and the area of the spots  is progressively reduced.
This gives rise to systematically negative residuals between the observations and the best-fit model when
$\eta > 0$. The optimal value of $\eta$ depends on the information content of the light curve, which in turn depends on the ratio of the amplitude of its rotational modulation to the average standard deviation of its  points. In the case of CoRoT-7, the  amplitude of the  rotational modulation is $\sim 0.018$, while the average standard deviation of the  points is $\sim 2.7 \times 10^{-4}$ in relative flux units, giving a signal-to-noise ratio of $\sim 65-70$. This is {sufficient} to adopt the same regularization criterion applied in the case of CoRoT-2 \citep[see ][]{Lanzaetal09a}, {iterating the value of $\eta$ until the condition $|\mu_{\rm reg}| \simeq \epsilon_{0}$ is met, where $\mu_{\rm reg}$ is the mean of the residuals and $\epsilon_{0} \equiv \sigma_{0}/\sqrt{N}$ their standard error, i.e., the ratio of their standard deviation $\sigma_{0}$ in the case of the unregularized best fit (i.e., for $\eta=0$) to the square root of the number $N$ of  points in each fitted subset of the light curve having a duration  $\Delta t_{\rm f}$ (see below).} 

In the case of the Sun, by assuming a fixed distribution of the filling factor, it is possible to obtain a good fit of the irradiance changes only for a limited time interval $\Delta t_{\rm f}$, not exceeding 14 days, which is the lifetime of the largest sunspot groups dominating the irradiance variation. In the case of other active stars, the value of        $\Delta t_{\rm f}$ must be determined from the observations themselves, looking for the {longest time interval} that allows  a good fit with the applied model (see Sect.~\ref{model_param}). 

The optimal values of the spot and facular contrasts and of the facular-to-spotted area ratio $Q$ in stellar active regions are  unknown a priori. In our model the facular contrast $c_{\rm f}$ and the parameter $Q$ enter as the product $c_{\rm f} Q$, so we can fix $c_{\rm f}$ and vary $Q$, estimating its best value 
 by minimizing the $\chi^{2}$ of the model, as shown in Sect.~\ref{model_param}. Since the number of free parameters of the ME model is large, for this specific application we make use of the model of \citet{Lanzaetal03}, which fits the light curve by assuming only three active regions to model the rotational modulation of the flux plus a uniformly distributed background to account for the variations of  the mean light level. This procedure is the same {that was used to fix} the value of $Q$ in the cases of CoRoT-2 and CoRoT-4 
\citep[cf. ][]{Lanzaetal09a,Lanzaetal09b}.  

We shall assume an inclination of the rotation axis of CoRoT-7 of $ i \simeq 80^{\circ}$ (see Sect.~\ref{model_param}). Since the information on spot latitude that can be extracted from the rotational modulation of the flux for such a high inclination is negligible, the ME regularization virtually puts all the spots at the sub-observer latitude (i.e., $90^{\circ} -i \approx 10^{\circ}$) to minimize their area and maximize the entropy. Therefore, we are limited to mapping  only the distribution of the active regions vs. longitude, which can be done with a resolution {better than}   $\sim 50^{\circ}$ \citep[cf. ][]{Lanzaetal07,Lanzaetal09a}. Our ignorance of the true facular contribution  may lead  to systematic errors in the active region longitudes derived by our model {because faculae produce an increase of the flux when they are close to the limb, leading to a systematic shift of the longitudes of the active regions used to reproduce the observed flux modulation, }  as discussed  by \citet{Lanzaetal07} {for} the case of the Sun {and illustrated by \citet[][ cf.~ Figs.~4 and 5]{Lanzaetal09a} for CoRoT-2.} 

\subsection{Activity-induced radial velocity variations}
\label{line_prof_model}

Surface magnetic fields in late-type stars produce brightness and convection inhomogeneities that shift and distort their spectral line profiles leading to apparent RV variations 
{\citep[cf., e.g., ][]{SaarDonahue97,Saaretal98,Huberetal09}}. 
To compute the apparent RV variations induced by stellar active regions, we adopt a simple model for each line profile considering the residual profile $R(\lambda)$ at a wavelength $\lambda$ along the line, i.e.: $R(\lambda) \equiv [1 - I(\lambda)/I_{\rm c}]$, where $I(\lambda)$ is the specific intensity at wavelength $\lambda$ and $I_{\rm c}$ the intensity in the  continuum adjacent to the line. The local residual profile is assumed to have a Gaussian shape with thermal and microturbulent width 
$\Delta \lambda_{\rm D}$ \citep[cf., e.g., ][ Ch.~1]{Gray88}, i.e.:
\begin{equation}
R(\lambda) \propto \exp \left[- \left( \frac{\lambda -\lambda_{0}}{\Delta \lambda_{\rm D}} \right)^{2} \right],
\label{local_prof}
\end{equation} 
where $\lambda_{0}$ is the local central wavelength. For simplicity, $R(\lambda)$ is assumed to be independent of the disc position and the presence of spots or faculae, so the local specific intensity along the line $I(\lambda)$ depends only on the variation of the continuum intensity $I_{\rm c}$ owing to limb-darkening and the effects of dark spots and bright faculae, as specified by  Eq.~(\ref{delta_flux}). 

To include the effects of surface magnetic fields on convective motions, we consider 
the decrease of macroturbulence velocity  and the reduction of convective blueshifts in active regions.

Specifically, we assume a local radial-tangential macroturbulence, as introduced by \citet{Gray88}, with a distribution function $\Theta$ for the radial velocity $v$ of the form: 
\begin{equation}
\Theta (v, \mu) = \frac{1}{2} \frac{\exp \left[- \left( \frac{v}{\zeta_{\rm RT} \sqrt{1 -\mu^{2}}} \right)^{2} \right]}{\sqrt{\pi} \zeta_{\rm RT} \sqrt{1 - \mu^{2}}} + \frac{1}{2} \frac{\exp \left[- \left( \frac{v}{\zeta_{\rm RT} \mu} \right)^{2} \right]}{\sqrt{\pi} \zeta_{\rm RT} \mu}, 
\label{macroturb}
\end{equation}
where $\zeta_{\rm RT}$ is the macroturbulence and $\mu$ is given by Eq.~(\ref{mu}). The  profile emerging from each surface element is obtained by convolving the local profile in Eq.~(\ref{local_prof}) with the  Doppler shift distribution as generated by the macroturbulence function in Eq.~(\ref{macroturb}). 
{The parameter $\zeta_{\rm RT}$ is assumed to be  reduced in spotted and facular areas, according to their total filling factor, i.e., $\zeta_{\rm RT} = \zeta_{0} [1- (Q+1)f]$, where $\zeta_{0}$ is the unperturbed macroturbulence.}  This is a convenient parameterization of the quenching effect of surface magnetic fields  on convective turbulence, at least in the framework of our simplified model, {and is supported by the observations of the reduction of turbulent velocity fields in the plage regions of the Sun \citep[cf., e.g., ][]{Titleetal92}. } 

{
Convective blueshifts arise from the fact that in stellar photospheres most of the flux comes from the extended updrafts {at the centre of}  convective granules. At the centre of the stellar disc, 
the vertical component of the convective velocity  produces a maximum blueshift, while
the effect vanishes at the limb where the projected  velocity is zero. 
The cores of the deepest spectral lines form in the upper layers of the photosphere where the vertical convective velocity is low, while the cores of shallow lines form in deeper layers with a greater  vertical velocity. Therefore, the cores of shallow lines are blueshifted with respect to the cores of the deepest lines. \citet{Gray09} showed this effect by plotting the endpoints of line bisectors of shallow and deep lines on the same velocity scale. He showed that the amplitude of the relative blueshifts scales with the spectral type and the luminosity class of the star. For the G8V star $\tau$ Ceti, {which} has a spectral type close to the G9V of CoRoT-7, the convective blueshifts {should be}  similar to those of the Sun, so we adopt solar values in our simulations. 
In active regions, vertical convective motions are quenched, so we observe an apparent redshift of the spectral lines in spotted and facular areas in comparison to the unperturbed photosphere. \citet{Meunieretal10} have quantified {this effect} in the case of the Sun and we adopt their results, considering an apparent redshift $\Delta V_{\rm f}=200$ m~s$^{-1}$ in faculae and $\Delta V_{\rm s} = 300$ m~s$^{-1}$ in cool spots. 

In principle,  the integrated effect of convective redshifts can be measured  in a star by comparing RV measurements obtained with two different line masks, one including the shallow lines and the other the  deep lines \citep[cf. ][]{Meunieretal10}. In the case of CoRoT-7, {lacking  such  measurements}, we apply the results of  \citet{Gray09} and adopt solar-like values as the best approximation. 

Considering solar convection as a template, intense downdrafts are localized in the dark lanes at the boundaries of the upwelling granules, but they contribute a significantly smaller flux because of their lower brightness and smaller area. {While a consideration of those downdrafts is needed to simulate the shapes of line bisectors, it is beyond our simplified approach that assumes that the whole profile of our template line forms at the same depth inside the photosphere.
Therefore,  we restrict our model to the mean apparent RV variations,}  neglecting the associated variations of the bisector shape and do not include the effect of convective downdrafts  as well as those of other surface flows typical of solar active regions, such as the Evershed effect in sunspots \citep[cf., e.g., ][]{Meunieretal10}.   

The local central wavelength $\lambda_{0}$ of the $k$-th surface element at time $t_{j}$ is given by: 
$\lambda_{0kj} = \lambda_{\rm R} (1+ v_{kj}/c)$, where $v_{kj}$ is its radial velocity: 
\begin{equation}
v_{kj} = - (v \sin i) \sin \theta_{k} \sin [\ell_{k} + \Omega  (t_{j} - t_{0})] + \Delta V_{\rm cs},
\end{equation}
with $c$ the speed of light, 
 $v \sin i$  the stellar projected rotational velocity,  $\lambda_{\rm R}$ the rest wavelength of the spectral line, and $\Delta V_{\rm cs}$ the apparent convective redshift arising from the reduction of convective blueshifts in spots and faculae, which is parameterized as:
\begin{equation}
\Delta V_{\rm cs} = ( \Delta V_{\rm s}  + Q  \Delta V_{\rm f}) f_{k} \mu_{kj},  
\end{equation}
where $f_{k}$ is the spot filling factor of the element, $Q$ the facular-to-spotted area ratio, and $\mu_{kj}$ is given by Eq.~(\ref{mu}). }

We integrate the line specific intensity at a given wavelength over the disc of the star using a subdivision into  $1^{\circ} \times 1^{\circ}$ surface elements to obtain the flux along the line profile $F(\lambda, t_{j})$ at a given time $t_{j}$. To find the apparent stellar RV, we can fit a Gaussian to the  line profile $F(\lambda, t_{j})$, or we can determine the centroid of the profile as:
\begin{equation}
 \lambda_{\rm ce} (t_{j}) = \frac{\int \lambda {\cal R}(\lambda, t_{j})\, d \lambda}{\int {\cal R}(\lambda, t_{j})\, d \lambda}, 
\label{centroid}
\end{equation}
where ${\cal R} \equiv [1-F(\lambda, t_{j})/F_{\rm c} (t_{j})]$ is the residual profile of the spectral line computed from the ratio of the flux in the line $F(\lambda, t_{j})$ to the flux $F_{\rm c} (t_{j}) $ in the adjacent continuum at any given time $t_{j}$. 

 A single line profile computed with the above model can be regarded as a cross-correlation function ({hereafter} CCF) obtained by cross-correlating the whole stellar spectrum with a line mask consisting of Dirac delta functions  giving the rest wavelength and depth of each individual line \citep[e.g., ][]{Quelozetal01}. Therefore,  to derive the RV from a single synthetic line profile is equivalent to measure the RV from a  CCF, either by fitting it with a Gaussian, or by computing its centroid wavelength.  A better approach would be that of simulating the whole stellar optical spectrum or an extended section of it in order to account for the wavelength  dependence of the spot and facular contrasts as well as of the convective inhomogeneities \citep[cf., e.g.,  ][]{Desortetal07,Lagrangeetal10,Meunieretal10}. Again, in view of our limited knowledge of the distribution of active regions on the stellar surface, our simplified approach is adequate to estimate the RV perturbations induced by magnetic fields. 

\subsubsection{Some illustrative examples}
\label{illustrative_examples}

\begin{figure}[]
\centerline{
\includegraphics[width=8cm,height=12cm]{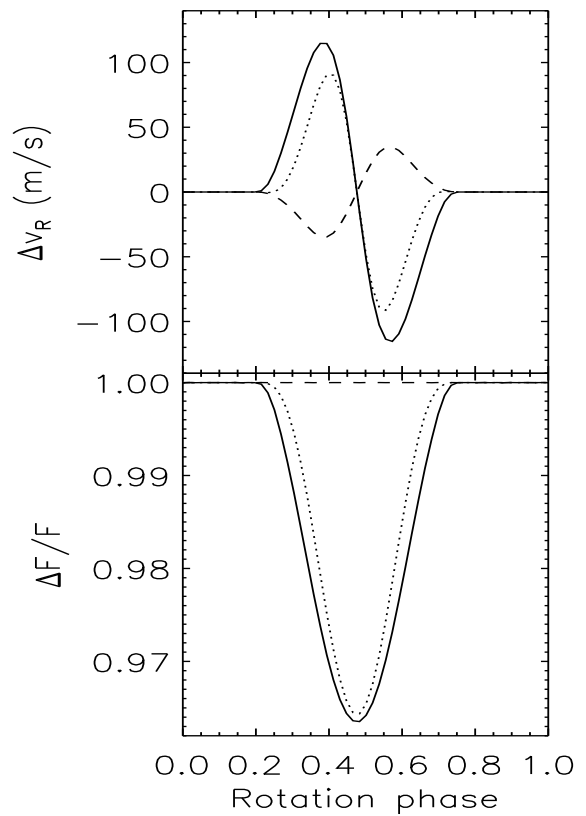}} 
\caption{{\it Upper panel:} {Apparent RV variation (upper panel) and relative stellar flux (lower panel) vs. the rotation phase produced by a single active region on a star seen equator-on. Different linestyles indicate different kinds of inhomogeneities inside the active region. Dashed: reduced macroturbulence without brightness perturbation; solid: dark spot and reduced macroturbulence; dotted: dark spot, bright facula, and reduced macroturbulence.  The flux is measured in units of the unperturbed stellar flux. Note that  macroturbulence perturbation alone does not affect the variation of the stellar flux.} }
\label{rv_perturb}
\end{figure}

To illustrate how active regions affect the measurement of  stellar RV, we consider the simple case of a single active region on a slowly rotating star, assuming stellar parameters similar to the cases presented by \citet{SaarDonahue97} and \citet{Desortetal07} for comparison purposes.  Specifically, we adopt 
$v \sin i =6.5$~km~s$^{-1}$ and a macroturbulence $\zeta_{0} = 5.5$~km~s$^{-1}$ to make macroturbulence effects clearly visible. 
{ For the moment, we do not include convective redshifts in spots and faculae because they were not taken into account in those simulations.} 
We consider a single active region of {square shape} $18^{\circ} \times 18^{\circ}$ located at the equator of a star observed equator-on, to maximize the RV variation. The spot filling factor is assumed to be $f=0.99$ and their contrast $c_{\rm s}=0.01$, i.e., the spot effective temperature is much lower than the photospheric temperature. When faculae are included, we adopt $c_{\rm f}=0.115$ and $Q=9$ to make their contribution clearly evident. In Fig.~\ref{rv_perturb} we show the apparent RV  and the relative flux variations produced by the active region vs. the rotation phase. The RV variations are computed by fitting a Gaussian to the line profile, as in \citet{SaarDonahue97}, while by applying the centroid method we find values which differ by at most $10-15$ percent. 
The greatest apparent variation is obtained in the case of a purely dark spot without any bright facula or reduced macroturbulence. In this case, the line profile integrated over the stellar disc shows a bump at the RV of the spot because the intensity of the local continuum is reduced in the spot \citep[see Fig.~1 in ][ for an illustration of the effect]{VogtPenrod83}. A comparison with the  amplitudes calculated with the formulae of \citet{SaarDonahue97} or \citet{Desortetal07} shows agreement within $\sim 10$ percent. 
{Note that the RV perturbation is positive and the  flux is decreasing when the spot is on the approaching half of the disc. When the spot transits across the central meridian,  the RV perturbation becomes zero and the flux reaches  a relative minimum. Finally, the RV perturbation becomes negative when the spot is on the receding half of the disc and the flux is increasing. } 

The inclusion of the radial-tangential macroturbulence
produces a decrease of the RV amplitude because macroturbulence is reduced in spots so that the local line profile becomes narrower and deeper while its equivalent width is not significantly affected \citep[][]{Gray88}. This reduces the height of the bump in the integrated profile acting in the opposite direction to the perturbation produced by a dark spot, as it is  evident in the case in which only a macroturbulence reduction is included in the model. Note that the wide-band flux variations are not affected by the inclusion of the macroturbulence in any case. 

\begin{figure}[]
\centerline{
\includegraphics[width=8cm,height=12cm]{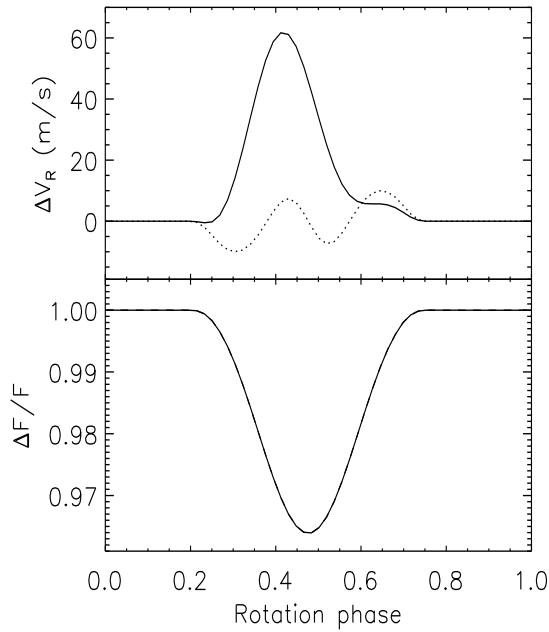}} 
\caption{{Same as Fig.~\ref{rv_perturb} when including (solid line) and not including (dotted line) the convective redshifts in cool spots and bright faculae in the simulation (see the text for the active region parameters). Note that the redshifts do not affect the stellar flux, so both cases coincide in the lower panel. } }
\label{conv_shifts} 
\end{figure}

 The effect of the faculae is to increase the local continuum intensity when an active region is close to the limb, leading to a greater local absorption which produces a dip on the integrated line profile. Since the contrast of the faculae is greatly reduced when they move close to the centre of the disc, this has the effect of narrowing the RV peak produced by an active region when dark spots and faculae are simultaneously present, i.e., faculae increase the  frequency of the RV variation. Specifically, the frequency of the RV variation in the case of a single active region is twice the rotation frequency when dark spots or a reduction of the macroturbulence are present {(cf. Fig.~\ref{rv_perturb}, upper panel, where we see a complete RV oscillation along one transit of the active region that lasts  half a stellar rotation)}.  
The inclusion of solar-like faculae can increase the frequency of the variation up to four times the rotation frequency when facular effects become important  (i.e., $c_{\rm f} Q > 1$). {This is illustrated in the upper panel of Fig.~\ref{conv_shifts} where the dotted line displays the RV variation produced by an active region with a sizeable facular contribution (see below) showing  two complete oscillations along one transit.  }

Next we consider the  apparent redshifts owing to the reduction of convective blueshifts in spots and faculae. To make the effect clearly evident, now we adopt a spot contrast $c_{\rm s}=0.665$, a facular contrast $c_{\rm f}=0.115$, and $Q = 4.5$. For the convective redshifts, we assume:
$\Delta V_{\rm s} = 500$ m~s$^{-1}$ and $\Delta V_{\rm f}= 300$ m~s$^{-1}$. 
A smaller spot contrast, e.g., $c_{\rm s} =0.01$, as in the previous simulations, {makes the contribution of the local redshifted line profile coming from the spotted area very small, reducing the corresponding RV  perturbation.}  In Fig.~\ref{conv_shifts}, we plot in the upper panel the RV variation produced by the single  active region previously considered, with all the other parameters kept at the previous values (solid line). For the purpose of comparison, we also plot  the RV perturbation obtained without convective redshifts (dotted line). In this case, the effects of cool spots and bright faculae almost balance each other because of the higher spot contrast, and the oscillation of the RV shows a high frequency. On the other hand, the inclusion of convective redshifts {brings} the frequency of the RV variation close to the rotation frequency, with a single maximum per rotation. This happens because the perturbation is always positive (cf. middle panel in Fig.~\ref{conv_shifts}), while all the other effects change  their sign when the active region transits from the approaching to the receding half of the stellar disc. Moreover, the convective redshifts do not depend per se on the contrast factors $c_{\rm s}$ and $c_{\rm f}$ and the shift due to the faculae is amplified by the  facular-to-spotted area ratio $Q$. The modulation of the continuum flux is of course not affected by the convective shifts, so we obtain the same light curve {independent of} the value of $\Delta V_{\rm s}$ and $\Delta V_{\rm f}$ (cf. Fig.~\ref{conv_shifts}, lower panel). 

Therefore, {particularly for not too dark  spots (i.e., $c_{\rm s} \ga 0.4$) and small $v \sin i$,} convective redshifts cannot be neglected in the simulation of the activity-induced RV perturbations in solar-like stars, as pointed out by \citet{Meunieretal10}. In the case of the Sun, their inclusion not only increases the RV perturbation up to an order of magnitude, but also {brings its dominant period close to the solar rotation period, while without their  inclusion they find that the dominant period  corresponds to the first harmonic of the solar rotation period. }

We note that the amplitude of the RV variation produced by an active region depends on several parameters that are poorly known, i.e., the latitude of the active region, the spot and facular contrasts, the convective redshifts, and the macroturbulence parameter which is difficult to  disentangle from rotational broadening in a slowly rotating star such as CoRoT-7 \citep[][]{Legeretal09}. Moreover, the spot and facular contrasts depend on the wavelength \citep{Lanzaetal04}, leading to a difference of $\approx 10$ percent in the RV variations as derived from different orders of an echelle spectrum \citep[cf., e.g., ][]{Desortetal07} and the convective redshift depends on the formation depth of a spectral line. 

The simultaneous presence of several active regions gives rise to a complex line profile distortion in the case of a slowly rotating star because {the perturbations of different  regions overlap in the line profile due to the weak rotational broadening.}  This implies an additional $15-20$ percent uncertainty in the determination of the central wavelength of the profile by the Gaussian fit or the centroid method, {as found by comparing the RV determinations obtained with the two methods in the case of line profiles simulated with infinite signal-to-noise ratio and perfectly regular sampling.}  In consideration of all these limitations, the absolute values of the RV variations computed with our model are uncertain by $20-30$ percent in the case of complex distributions of active regions, such as those derived by our spot modelling technique as applied to CoRoT-7 light curves.      

\subsubsection{Radial velocity variations from spot modelling}
\label{rv_from_model}

{
We can use the distribution of active regions  as derived from our  spot modelling to synthesize the corresponding RV variations according to the approach outlined in Sect.~\ref{line_prof_model}. Since our spot model assumes that active regions are stable for the time interval of each fitted time series $\Delta t_{\rm f}$, the distribution of surface inhomogeneities  can be used to synthesize the RV variations having a timescale of $\Delta t_{\rm f}$ or longer. Active regions with a shorter lifetime produce a photometric modulation which appears in the residuals of the best fit to the light curve. As we shall see in Sect.~\ref{light_curve_model}, most of the short-term variability occurs on time scales of $4-5$ days, i.e., significantly shorter than the rotation period of CoRoT-7, so we can neglect, as a first approximation, the variation of the disc position of those active regions due to stellar rotation and {estimate their area} as if they were located at the centre of the disc.
{Specifically,  first we express the residuals as the relative deviation $\Delta F/F_{0}$ between the observed flux and its best fit measured in units of the reference unspotted flux $F_{0}$; then we subtract from $\Delta F/F_{0}$ its mean value $\mu_{\rm res} \equiv \langle \Delta F /F_{0} \rangle$ which corresponds to a uniformly distributed population of active regions that do not produce any RV variation. 
At each given time $t$,
we adopt the relative deviation $| \Delta F/F_{0} - \mu_{\rm res}| $ as a measure of the filling factor of the  active regions producing the short-term RV variations. In doing so, we neglect limb-darkening effects and assume that those active regions  consist of completely dark spots ($c_{\rm s}=0$). Finally, we compute the RV perturbation due to such brightness inhomogeneities by means of Eq.~(1) of \citet{Desortetal07} obtaining: }
\begin{equation}
\Delta V_{\rm Ri} = 800 (\frac{\Delta F}{F} - \mu_{\rm res}) (v \sin i)^{1.1},
\label{rv_short_term}
\end{equation}
where the RV perturbation is measured in m~s$^{-1}$ and the $v \sin i$ of the star in km~s$^{-1}$; for simplicity, Eq.~(\ref{rv_short_term}) assumes that the perturbation has the {same sign as} $\Delta F/F_{0} - \mu_{\rm res}$ which should produce statistically correct results in the case of a sufficiently long time series, i.e.,  with equally probable positive and negative deviations. {In our case, Eq.~(\ref{rv_short_term}) gives an upper limit for the amplitude of the RV perturbation because it assumes that an active region spans the diameter of the stellar disc. This is appropriate because we want to assess the impact of short-lived active regions on the detection of  exoplanets. }
 On the other hand, the effects of  convective redshifts in short-lived active regions is estimated as:
\begin{equation}
\Delta V_{\rm Rc} = (\Delta V_{\rm s } + Q \Delta V_{\rm f}) (\frac{\Delta F}{F} - \mu_{\rm res}), 
\label{rv_short_csh}
\end{equation}   
where the  sign of $\Delta V_{\rm Rc}$ has been chosen to maximize the total short-term RV perturbation when added to that of Eq.~(\ref{rv_short_term}). 
By adding both kinds of short-term perturbations to the long-term RV modulation produced by the  active regions mapped by the spot modelling, we compute the total synthetic RV variation induced by stellar activity. 
}

\section{Model parameters}
\label{model_param}

The fundamental stellar parameters are taken from \citet{Legeretal09} and are listed in 
Table~\ref{model_param_table}.  {The limb-darkening parameters $a_{\rm p}$, $b_{\rm p}$, and $c_{\rm p}$ (cf. Sect.~\ref{spotmodel}) } have been derived from \citet{Kurucz00} model atmospheres for $T_{\rm eff} = 5275$~K, $\log g = 4.50$~(cm s$^{-2}$) and solar abundances, by adopting the CoRoT white-band transmission profile given by  \citet{Auvergneetal09}. { Recently, \citet{Brunttetal10} have  slightly revised stellar parameters, but adopting their values does not produce any significant modification as far as our modelling is concerned.}
The rotation period adopted for our spot modelling  has been derived from a periodogram analysis of the light curve giving $P_{\rm rot} = 23.64 \pm 3.62$ days. { The uncertainty of this period determination is derived from the total extension of the time series and represents an upper limit. As we shall see below, our spot model shows that the starspots have a remarkable differential rotation which contributes to an uncertainty  of the stellar rotation period of $\sim 6$ percent, i.e., of $\pm 1.42$ days (cf. Sect.~\ref{spot_model_res}). } 

\begin{table}
\noindent 
\caption{Parameters adopted for the light curve and RV modelling of CoRoT-7.}
\begin{tabular}{lrr}
\hline
 & & \\
Parameter &  & Ref.$^{a}$\\
 & & \\ 
\hline
 & &  \\
Star Mass ($M_{\odot}$) & 0.93 & L09  \\
Star Radius ($R_{\odot}$) & 0.87 & L09  \\
$T_{\rm eff}$ (K) & 5275 &  L09 \\
$\log g$ (cm s$^{-2}$) & 4.50 & L09 \\ 
$a_{\rm p}$ & 0.284 & La10 \\
$b_{\rm p}$ & 0.961 & La10 \\
$c_{\rm p}$ & -0.254 & La10 \\ 
$P_{\rm rot}$ (days) & 23.64 & La10 \\
$\epsilon$ & $8.54 \times 10^{-6}$ & La10 \\ 
Inclination (deg) & 80.1 & L09  \\
$c_{\rm s}$  & 0.577 & La10 \\
$c_{\rm f}$  & 0.115 & L04 \\ 
$Q$ & 1.0  & La10 \\ 
$\Delta t_{\rm f}$ (d) & 18.688, 14.532 & La10 \\ 
$v \sin i$ (km~s$^{-1}$) & 1.7 & Br10 \\
$\Delta \lambda_{\rm D} $ (km s$^{-1}$) &  2.33 & La10 \\
$\zeta_{0}$ (km~s$^{-1}$) & 1.2 & Br10 \\  
$\Delta V_{\rm s}$ (m~s$^{-1}$) & 300 & G09, Me10 \\
$\Delta V_{\rm f}$ (m~s$^{-1}$) & 200 & G09, Me10 \\
& &   \\
\hline
\label{model_param_table}
\end{tabular}
~\\
$^{a}$ References: Br10: \citet{Brunttetal10}; G09: \citet{Gray09}; L04: \citet{Lanzaetal04}; L09: \citet{Legeretal09}; La10: present study; Me10: \citet{Meunieretal10}.
\end{table}

The polar flattening of the star due to the centrifugal potential  is computed in the Roche approximation with a rotation period of 23.64 days. {The relative difference between the equatorial and the polar radii  $\epsilon$ is } $8.54 \times 10^{-6}$ which induces a completely negligible relative flux variation of $\approx 10^{-7}$  for a spot coverage of $\sim 2$ percent as a consequence of the gravity darkening of the equatorial region of the star. 

The inclination of the stellar rotation axis is impossible to constrain through the observation of the Rossiter-McLaughlin effect due to the very small planet relative radius $R_{\rm p}/R_{\rm star} = 0.0187 \pm 3.0 \times 10^{-4}$, the small rotational broadening of the star $v \sin i < 3.5$ km~s$^{-1}$,  and its intrinsic line profile variations due to stellar activity \citep{Quelozetal09}. Nevertheless, we assume that the stellar rotation axis is normal to the orbital plane of the transiting planet, i.e., with an inclination of $80\, \fdg 1 \pm 0\, \fdg 3$ from the line of sight \citep[cf. ][]{Legeretal09}. 

The maximum time interval that our model can accurately fit with a fixed distribution of active regions $\Delta t_{\rm f}$ has been determined by dividing the total interval, $T= 130.8152$ days, into $N_{\rm f}$ equal segments, i.e., $\Delta t_{\rm f} = T/N_{\rm f}$, and looking for the minimum value of $N_{\rm f}$ that allows us a good fit of the light curve, as measured by the $\chi^{2}$ statistics. We found that for $N_{\rm f} < 7$ the quality of the best fit degrades significantly with respect to $N_{\rm f} \geq 7$, owing to a  substantial evolution of the pattern of surface brightness inhomogeneities. Therefore, we adopt   $\Delta t_{\rm f} = 18.688$ days as the maximum time interval to be fitted with a fixed distribution of surface active regions in order to  estimate the best value of the parameter $Q$ (see below). This confirms the result of \citet{Quelozetal09} who estimated a starspot coherence time of $\sim 18$ days. However, for the spot modelling in Sects.~\ref{light_curve_model} and ~\ref{spot_model_res}, we shall adopt $N_{\rm f} = 9$, corresponding to $\Delta t_{\rm f} = 14.532$ days, which provides a better time resolution to study the evolution of the spot pattern during the intervals with  faster modifications. 

To compute the spot contrast, we adopt the same mean temperature difference as derived for sunspot groups from their bolometric contrast, i.e., 560~K \citep{Chapmanetal94}. In other words, we assume a spot effective temperature of $ 4715$~K, yielding a contrast $c_{\rm s} = 0.577$ in the CoRoT passband \citep[cf. ][]{Lanzaetal07}.
A different spot contrast changes the absolute spot coverages, but does not significantly affect their longitudes and their time evolution, as discussed in detail by \citet{Lanzaetal09a}. The facular contrast is assumed to be solar-like with $c_{\rm f} = 0.115$ \citep{Lanzaetal04}. 

\begin{figure}[]
\centerline{
\includegraphics[width=8cm,height=6cm]{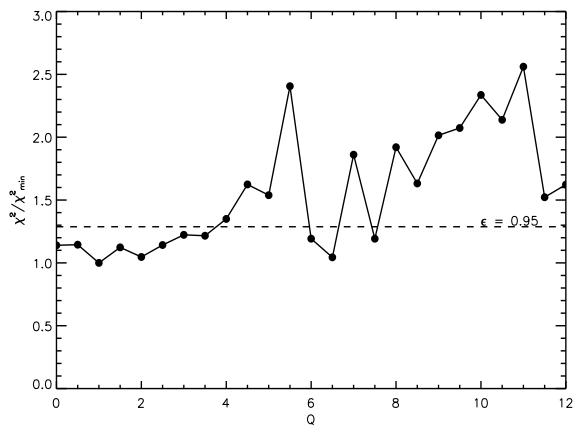}} 
\caption{
The ratio of the $\chi^{2}$ of the composite best fit of the entire time series  
to its minimum value vs. the parameter $Q$, i.e., the ratio of the {facular area  to  the cool spot area} in active regions. The horizontal dashed line indicates the 95 percent confidence level for $\chi^{2}/\chi_{\rm min}^{2}$, determining the interval of acceptable $Q$ values.
}
\label{qratio}
\end{figure}

The best value of the area ratio $Q$ of the faculae to the spots in each active region has been estimated by means of the model of \citet[][ cf. Sect.~\ref{spotmodel}]{Lanzaetal03}. In Fig.~\ref{qratio}, we plot the ratio $\chi^{2}/ \chi^{2}_{\rm min}$ of the 
total $\chi^{2}$ of the composite best fit of the entire time series to 
its minimum value $\chi^{2}_{\rm min}$, versus $Q$, and indicate the 95 percent confidence level as derived from the F-statistics \citep[e.g., ][]{Lamptonetal76}. Choosing $\Delta t_{\rm f} = 18.688$ days, we fit the rotational modulation of the active regions for the longest time interval during which they remain stable, modelling both the flux increase due to the facular component when an active region is close to the limb as well as the flux decrease due to the dark spots when the same region transits across the central meridian of the disc. In such a way, a measure of the relative facular and spot contributions can be obtained leading to a fairly accurate estimate of $Q$. 

The best value of $Q$ turns out to be $Q=1.0$, with an acceptable range extending from~$\sim 0$~to~$\sim 4$. There is also a small interval of formally acceptable values between $6.0$ and $6.5$, but we regard it as a   spurious outcome of the $\chi^{2}$ statistical fluctuations. { Although  a sizeable facular contribution cannot  be excluded on the basis of the photometric best fit, we shall find in  Sect.~\ref{results_rv} that $Q \leq 4$ is required to reproduce the amplitude of the observed RV variations.} In the case of the Sun, the best value of $Q$ is $9$ \citep{Lanzaetal07}. Thus our result indicates a lower relative contribution of the faculae to the  light variation  of CoRoT-7 than in the solar case. 
The amplitude of the rotational modulation of the star was $\sim 0.0185$ mag during CoRoT observations and $\sim 0.03$ mag during the campaign organized by \citet{Quelozetal09}, i.e., $\sim 6-10$ times that of the Sun at the maximum of the eleven-year cycle. This indicates that CoRoT-7 is more active than the Sun, which may account for the reduced facular contribution to its light variations, as suggested by \citet{Radicketal98} and \citet{Lockwoodetal07}. The ground-based photometry of \citet{Quelozetal09} also indicates that cool spots dominate the optical variability of the star since it becomes redder when it is fainter.  
{ The use of the chromatic information of the CoRoT light curves to estimate the spot and facular contrasts and filling factors is made impossible by our ignorance of the unperturbed stellar flux levels in the different colour channels which are needed to disentangle the flux perturbations due to spots and faculae, respectively. The continuous variations of the observed fluxes do not allow us to fix such reference levels so that we cannot unambiguously attribute a given flux modulation to cool spots or bright faculae. Moreover, the long-term drifts of the fluxes in the individual colour channels complicate the estimate of the flux variations in each of them, making this approach unfeasible.
}

To compute the RV variations induced by surface inhomogeneities, we assume a line rest wavelength of $600$~nm and a {local thermal plus microturbulence broadening $ \Delta \lambda_{\rm D} = 2.3$~km~s$^{-1}$. }  
{ The  $v \sin i = 1.7 \pm 0.2$~km~s$^{-1}$ is estimated from the stellar radius, inclination and rotation period following \citet{Brunttetal10}. The radial-tangential macroturbulence velocity is assumed $\zeta_{0} = 1.2 \pm 0.8$~km~s$^{-1}$ after \citet{Brunttetal10}.}
 Note that it is very difficult to obtain a good estimate of such a parameter because macroturbulence and  rotational broadening are largely degenerate owing to the slow rotation of the star \citep[cf. ][]{Legeretal09,Brunttetal10}. 
The convective redshifts in cool spots and faculae, $\Delta V_{\rm s}$ and $\Delta V_{\rm f}$, have been assumed to be similar to those of the Sun, following the discussion in Sect.~\ref{line_prof_model}.

\section{Results}
\label{results}

\subsection{Light curve model}
\label{light_curve_model}

\begin{figure*}[]
\centerline{
\includegraphics[width=8cm,height=16cm,angle=90]{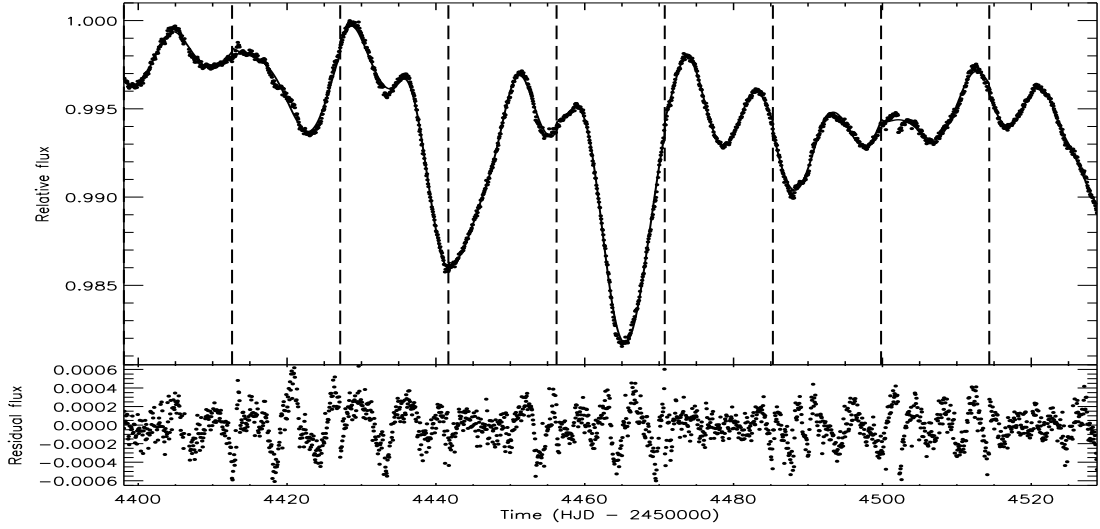}} 
\caption{{\it Upper panel:} {The out-of-transit light curve of CoRoT-7 (dots) and its  ME-regularized  best fit  for a facular-to-spotted area ratio of $Q=1.0$ (solid line). The flux is normalized to the maximum observed flux. The vertical dashed lines mark the individually fitted intervals of $\Delta t_{\rm f} = 14.532$ days. {\it Lower panel:} The corresponding residuals. }
}
\label{lc_bestfit}
\end{figure*}

We applied the model of Sect.~\ref{spotmodel} to the out-of-transit CoRoT-7 light curve, considering nine intervals of duration $\Delta t_{\rm f} = 14.532$ days. The best fit without regularization ($\eta = 0$) has a mean  $\mu_{\rm res} = 3.862 \times 10^{-6}$ and a standard deviation of the residuals $\sigma_{0} = 1.444 \times 10^{-4}$ in relative flux units. The Lagrangian multiplier $\eta$ is iteratively adjusted until the mean of the residuals $\mu_{\rm res} = -1.057 \times 10^{-5} \simeq - \sigma_{0} / \sqrt{N}$, where $N  =  196$ is the mean number of  points in each fitted light curve interval  $\Delta t_{\rm f}$. The standard deviation of the residuals of the regularized best fit is $\sigma = 1.951 \times 10^{-4}$. The composite best fit to the entire light curve is shown in the upper panel of Fig.~\ref{lc_bestfit} while the residuals are plotted in the lower panel. They show oscillations with a typical timescale of $\sim 4-5$ days that can be related to the rise and decay of small active regions that cover $\approx 0.1$ percent of the stellar disc, i.e., comparable to a {large} sunspot group.
These small active regions cannot be modelled by our approach because they do not produce a significant rotational flux modulation during the 23.6 days of the stellar rotation period as they move across the disc by only $\approx 60^{\circ}-75^{\circ}$ in longitude. By decreasing the degree of regularization, i.e., the value of $\eta$, we can {marginally} improve the best fit, but at the cost of introducing several small active regions that wax and wane from one time interval to the next and are badly constrained by the rotational modulation. Nevertheless, the oscillations of the residuals do not disappear completely even for $\eta = 0$, indicating that CoRoT-7 has a population of short-lived active regions with typical lifetimes of $4-5$ days. 

\begin{figure}[]
\centerline{
\includegraphics[width=8cm,height=8cm,angle=90]{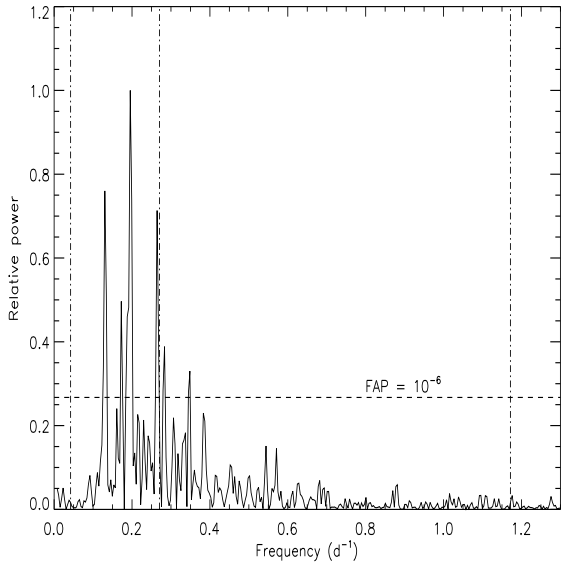}} 
\caption{{Lomb-Scargle periodogram of the residual time series  of Fig.~\ref{lc_bestfit}. The dashed horizontal line marks a false-alarm probability  of $10^{-6}$. The vertical dot-dashed lines mark the orbital frequencies of the planets CoRoT-7b (right, $P_{\rm orb}=0.8536$ days) and CoRoT-7c (middle, $P_{\rm orb}=3.698$ days) as well as  the stellar rotation frequency (left, $P_{\rm rot} = 23.64$ days). }
}
\label{res_period}
\end{figure}

A periodogram of the residuals is plotted in Fig.~\ref{res_period} and its  main peaks correspond to periods of 5.12 and 3.78 days, respectively, the latter close to the orbital period of CoRoT-7c {of 3.69 days}. Fitting a sinusoid {with this period,} we find a semiamplitude of the light variation of $6.47 \times 10^{-5}$ in relative flux units. The corresponding RV perturbation, estimated with Eqs.~(\ref{rv_short_term}) and (\ref{rv_short_csh}), has a semiamplitude of $\sim 0.4$ m~s$^{-1}$, {an order of magnitude smaller than the oscillation attributed to the planet CoRoT-7c whose semiamplitude is  $4.0 \pm 0.5$ m~s$^{-1}$. Concerning   CoRoT-7b,  the residual oscillations at its orbital frequency are practically zero. }


\subsection{Longitude distribution of active regions and stellar differential rotation}
\label{spot_model_res}

\begin{figure}[!h]
\centerline{
\includegraphics[width=7cm]{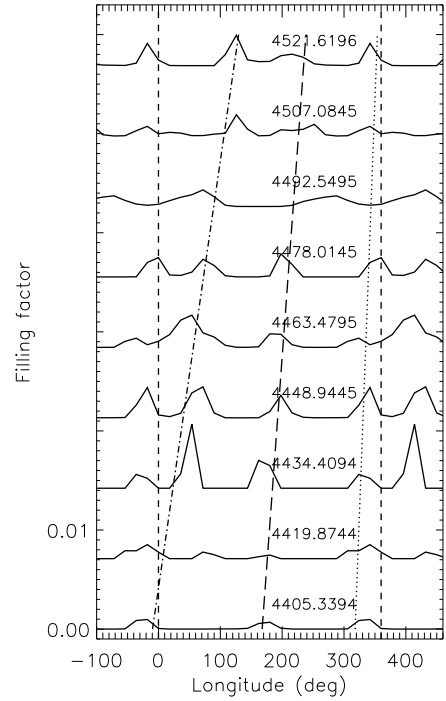}} 
\caption{{Distributions of the spotted area vs. longitude during the  time intervals centred on the labelled times (${\rm HJD}-2450000.0$), adopting $Q=1.0$. The plots have been vertically shifted to show the apparent  migration of individual active regions and repeated beyond $0^{\circ}$ and $360^{\circ}$.  The dotted, dot-dashed, and long-dashed lines trace the apparent migration of the three most conspicuous features, respectively (see  text for details). }
}
\label{long_distr}
\end{figure}

The distributions of the spotted area vs. longitude are plotted in Fig.~\ref{long_distr}
for the nine mean epochs of our individual subsets adopting a rotation period of 23.64 days. {Longitude zero corresponds to the sub-observer point  at the initial epoch, i.e., HJD~2454398.0719.} The longitude increases in the same direction as the stellar rotation. This is consistent with the reference frames adopted in our previous  studies \citep[][]{Lanzaetal09a,Lanzaetal09b}. {To allow a  comparison of the mapped active regions with the dips in the light curve, note that a feature at longitude $\ell$ crosses the central meridian of the star at  rotation phase $360^{\circ}- \ell$}.

{The spot distribution and its evolution suggest three large-scale active regions which show different 
intrinsic longitudinal motions. Assuming that this is due to a shear by differential rotation, we perform linear fits  by assuming  constant migration rates that are found to be} $0.31 \pm 0.12 $ deg/day for that fitted with a dotted line; $1.20 \pm 0.19$ deg/day for that fitted with a dot-dashed line; and $0.61 \pm 0.27$ deg/day for that fitted with a long-dashed line. The relative amplitude of the surface differential rotation, as estimated from the difference between the greatest and the lowest migration rates, is $ \Delta \Omega /\Omega = 0.058 \pm 0.017$; {this is only a lower limit since the spot latitudes are unknown.} Since CoRoT-7 is more active than the Sun, its active regions may cover a latitude range  greater than in the Sun where sunspot groups are confined to $\pm 40^{\circ}$ from the equator \citep[see, e.g., ][]{Strassmeier09}. 
For the Sun, one finds $ \Delta \Omega / \Omega \simeq 0.04-0.05$ by considering active regions confined to within  the sunspot belt, i.e., within $\pm \, (35^{\circ} - 40^{\circ})$ from the equator. Thus, CoRoT-7 may have a surface differential rotation comparable to the Sun if its active regions are also mainly localized at low latitudes.

It is interesting to compare the lower limit for the surface shear of CoRoT-7 with {the surface differential rotation estimated by \citet{Crolletal06} for the K2 dwarf $\epsilon$ Eridani. They found two spots rotating with periods of 11.35 and 11.55 days at  latitudes of $20.^{\circ}0$ and $31^{\circ}.5$, respectively, leading to a differential rotation amplitude of half of that of the Sun. This is consistent with a weak dependence of the absolute surface shear $\Delta \Omega$ on the angular velocity $\Omega$, because $\epsilon$ Eri has a rotation period roughly half of that of the Sun. Indeed, \citet{Barnesetal05} found $\Delta \Omega \propto \Omega^{0.15}$ for late-type stars at a fixed effective temperature. This suggests that the true pole-equator angular velocity difference in CoRoT-7 may be about a factor of $2-3$ greater than observed, if the active regions mapped by our technique are mainly localized at low latitudes. } 

\begin{figure}[!h]
\centerline{
\includegraphics[width=5cm,angle=90]{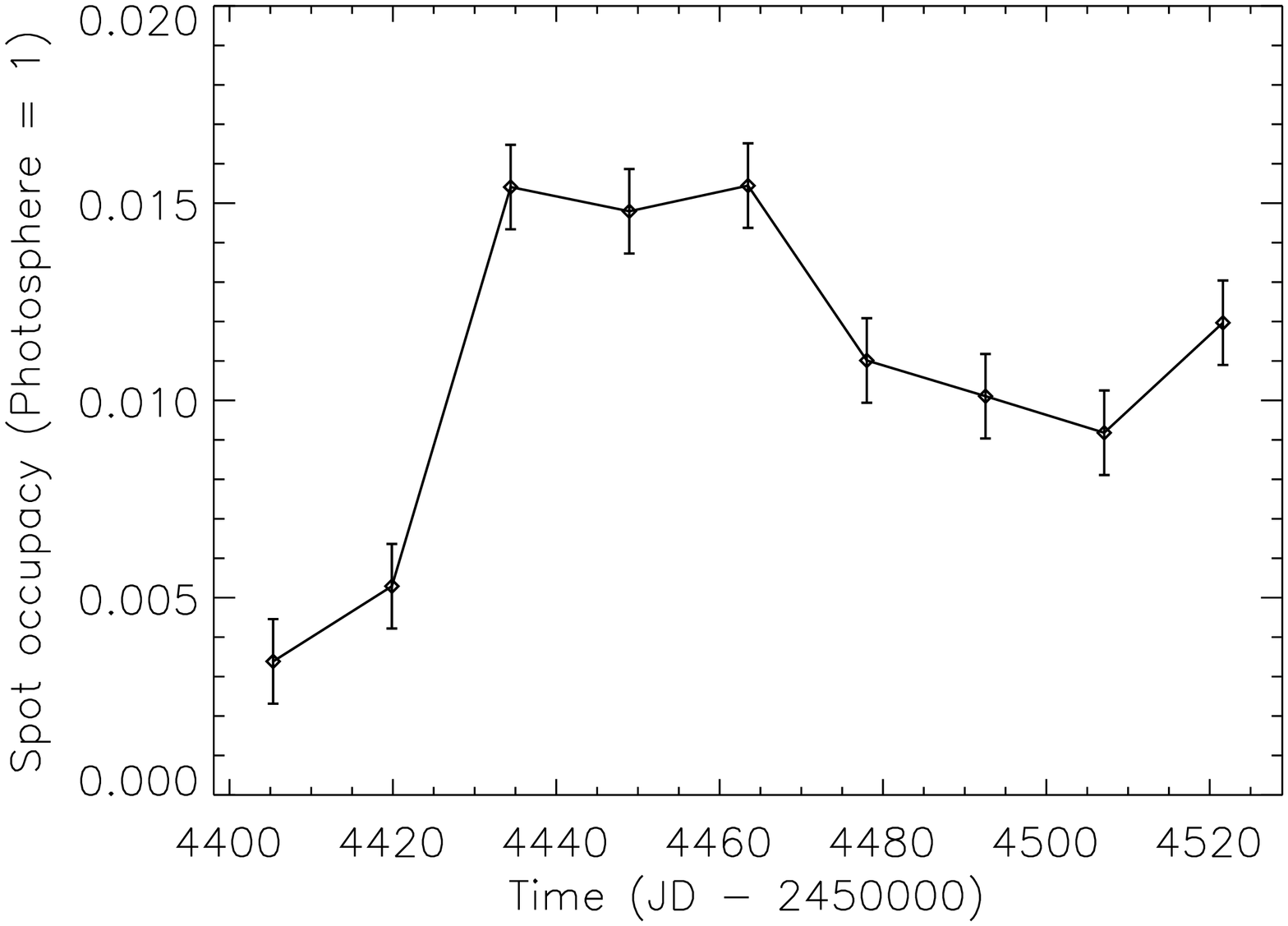}} 
\caption{{The total spotted area as derived from our lightcurve model illustrated in Figs.~\ref{lc_bestfit} and~\ref{long_distr}. The uncertainty of the area has been estimated from the errors of the photometric observations. }   
}
\label{total_area}
\end{figure}

The  stable active longitudes plotted in Fig.~\ref{long_distr} show significant area changes on a timescale as short as two weeks although their overall {lifetime} may exceed the duration of the light curve, i.e., 130 days. As a matter of fact, \citet{Quelozetal09} found that the rotational modulation of the optical flux observed during their ground-based campaign from December 2008 to February 2009 matched the extrapolation of the CoRoT light curve when a rotation period of 23.64 days was assumed, although 17 complete rotations had elapsed from CoRoT observations. This suggests that {the active longitudes may exist for several years, although some large active regions wax and wane on time scales ranging from two weeks (i.e., the time resolution of our mapping) to a few months.} The variation of the total spotted area {of our surface models} vs. time is plotted in Fig.~\ref{total_area} and shows a characteristic timescale  of $\sim 60-80$ days. Note that the absolute values of the area per longitude bin and of the total area depend on the spot and facular contrasts adopted for the modelling. Specifically, darker spots lead to a smaller total area while the effect of the facular contrast is more subtle and influences somewhat the longitudinal distribution of the active regions
{\citep[cf., e.g., ][ Figs. 4 and 5]{Lanzaetal09a}.  }

\subsection{Activity-induced radial velocity variations}
\label{results_rv}

\begin{figure*}[]
\centerline{
\includegraphics[width=16cm,height=10cm,angle=0]{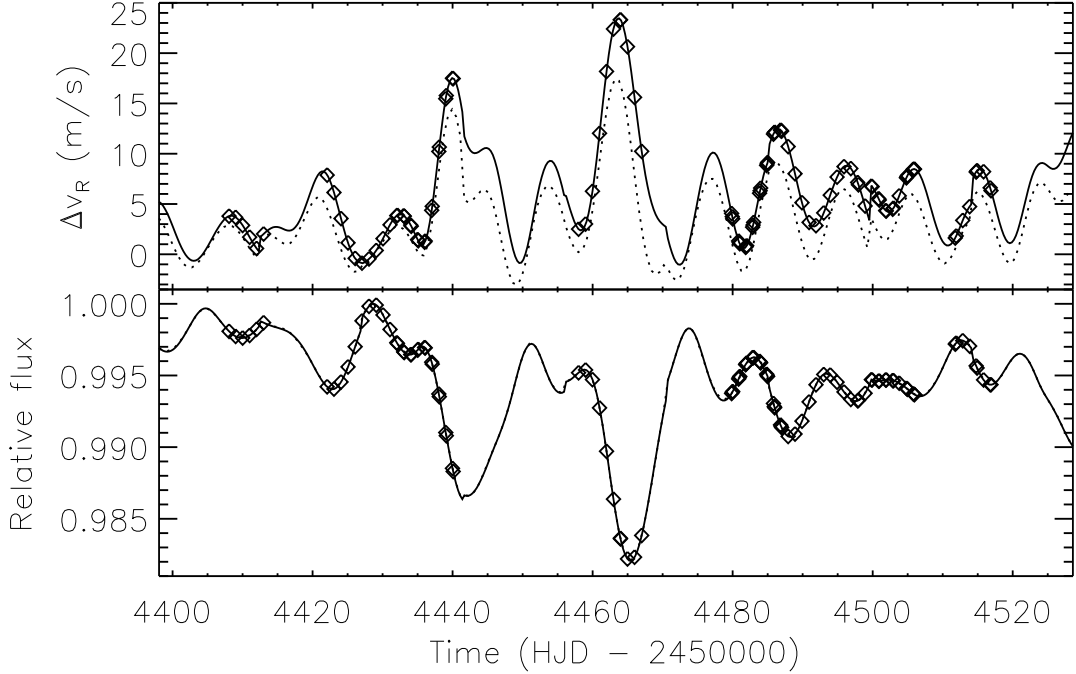}} 
\caption{{\it Upper panel:} The synthesized RV variation due to  stellar active regions vs. time as derived from our ME spot modelling {with $Q=1$ (solid line) and $Q=0$ (dotted line). To illustrate the effect of an exemplary ground-based sampling,} open diamonds mark the times of HARPS observations by \citet{Quelozetal09} after shifting their initial epoch to ten days after the beginning  of CoRoT observations. 
{\it Lower panel:} The synthesized relative flux variation in the CoRoT white passband vs. time (solid line) with  HARPS time sampling marked by  open diamonds. {Of course, the synthesized fluxes are the same both for models with ($Q=1$) and without ($Q=0$) faculae since they must reproduce the rotational modulation as observed by CoRoT. }
}
\label{rv_time}
\end{figure*}

To simulate the apparent RV changes induced by the distribution of active regions derived from our  light curve modelling, {we consider a spectral line with a rest wavelength of $600$ nm, i.e.,  close to the isophotal wavelength of CoRoT observations} for which our contrast coefficients are given. The spectral  resolution of the line profile is $\lambda/\Delta \lambda = 230,000$, and 50 profiles are computed per stellar rotation to warrant a good phase sampling. The RV is derived by computing the centroid of each  profile 
{which turns out to be more stable than fitting a Gaussian which sometimes leads to poor results in the presence of complex line profile distortions.}
First we consider the RV  perturbation produced by active regions with an evolutionary timescale equal to or greater than $\Delta t_{\rm f}= 14.532$ days as mapped by means of our spot modelling approach {for $Q=1$ and $Q=0$ (cool spots only)}. We plot the simulated RV variations vs. time in the upper panel of Fig.~\ref{rv_time} {(solid line for $Q=1$, dotted line for $Q=0$) that shows a peak-to-peak amplitude up to  $\sim 24.5$ m s$^{-1}$ for
 $Q=1$ and to $\sim 20.5$ m s$^{-1}$ for $Q=0$}.  Small discontinuities are seen at the endpoints of each $\Delta t_{\rm f}$ interval owing to the change of the spot configuration between  successive light curve fits. Their impact on the simulated RV time series is small and generally  comparable to the measurement errors (see below), so it is not necessary to correct for them. 

The observations by \citet{Quelozetal09} showed a peak-to-peak amplitude up to $\sim 40$ m s$^{-1}$ for the 95 percent of the measurements collected during their campaign from December 2008 to February 2009 while the photometric optical modulation was a factor of $\sim 1.6$ greater. Since the RV perturbation scales   approximately linearly with the spot filling factor, our computed amplitude {for $Q=1$} agrees well with the observed one, within $\approx 2-3$ percent, which is a good level of precision for this kind of simulations \citep[cf., e.g., ][]{SaarDonahue97,Desortetal07}. Note that most of the active regions mapped by  our spot modelling technique fall within $\pm 20^{\circ}$ from the equator, owing to the preference of ME regularization to put  spots close to the sub-observer latitude to minimize their area. 

The lower panel of Fig.~\ref{rv_time} shows the simultaneous CoRoT white band flux variation.  In the present case,  facular brightening and  macroturbulence reduction play only a minor role {in the wide-band flux and RV  variations, respectively,} because $Q=1.0$ and $\zeta_{0} = 1.2$~km~s$^{-1} < v \sin i$ (cf. Sect.~\ref{line_prof_model}). The dominant effect is the reduction of convective blueshifts in the spotted and facular areas.
{The diamonds show the sampling intervals of the RV observations  by \citet{Quelozetal09}.
Since their HARPS measurements were not simultaneous with CoRoT observations, the former have been arbitrarily shifted by ten days with respect to the latter to illustrate the effects of an examplary ground-based time sampling.}  

To simulate more realistic RV time series,  we add the short-term RV fluctuations associated with the residuals of our composite photometric best fit, computed according to the method outlined in Sect.~\ref{rv_from_model}. To account for HARPS measurement errors, we further add a Gaussian random noise with a standard deviation of 2.0~m~s$^{-1}$. {In such a way, we take into account both the RV variations produced by active regions with 
a lifetime comparable to $\Delta t_{\rm f}$ or longer, as mapped by our spot modelling approach, as well as the 
variations induced by active regions with a shorter lifetime whose photometric effects appear as the residuals 
of the best fits.} 

We compute $10^{5}$ RV time series for $Q=1$, differing by the realization of the Gaussian noise and the initial epoch which is chosen from a uniform random distribution to sample all the  parts of the synthetic  RV curve in Fig.~\ref{rv_time} and the short-term fluctuations. To allow for different initial epochs, we extend the synthetic  time series by mirror reflections at their endpoints. {We choose to compute $10^{5}$ simulations to have enough statistics to sample the distribution of the RV amplitudes down to a probability of the order of $10^{-4}$. }
\begin{figure*}[]
\centerline{
\includegraphics[width=16cm,height=10cm,angle=0]{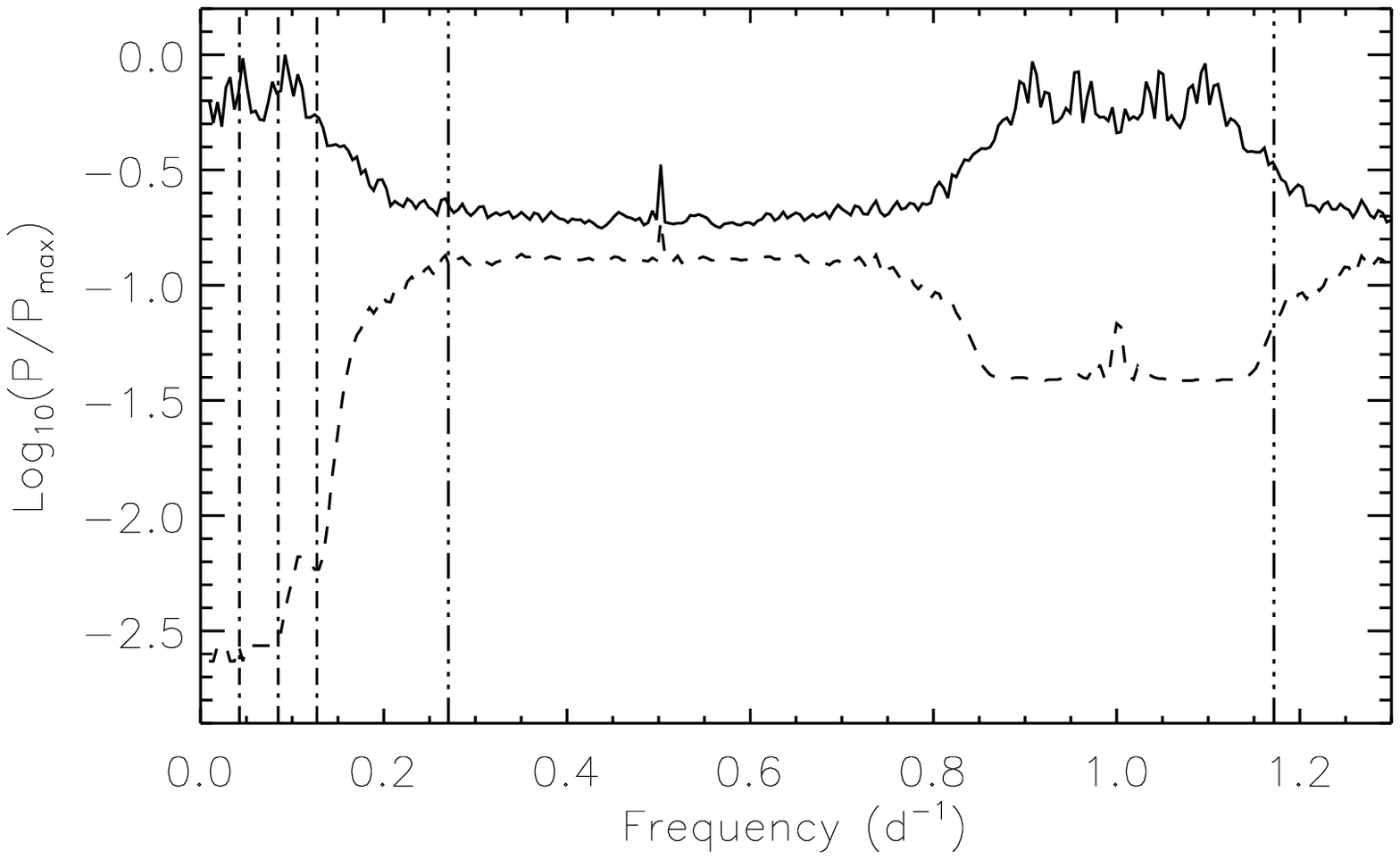}} 
\caption{ The average Lomb periodogram of  $10^{5}$ synthetic RV time series vs. frequency (solid line) and the average periodogram obtained after filtering the series with the three-harmonic model of \citet{Quelozetal09} (dashed line). The power is normalized to the maximum value of the unfiltered mean periodogram. The dot-dashed lines mark the rotation frequency, its first, and second harmonics, while the three-dot-dashed  lines mark the orbital frequencies of  CoRoT-7c (left) and CoRoT-7b (right), respectively. Note the  strong aliasing associated with the daily cadence of  the ground-based observations.  
}
\label{per_rv}
\end{figure*}

For each time series, we compute the Lomb periodogram which gives the squared amplitude of a best fitted sinusoidal signal as a function of the frequency for an uneven time sampling \citep{Lomb76}. 
On the other hand,  the amplitudes obtained with the Scargle periodogram are normalized in a different way,  as required for the computation of the false-alarm probability,  making the comparison of  the power level at different frequencies less immediate \citep{Scargle82}. 

{The power spectrum obtained by averaging the Lomb periodograms of our $10^{5}$ synthesized time series is plotted in the upper panel of Fig.~\ref{per_rv} (solid line). Strong aliases at $1.0$~day$^{-1}$ are clearly evident in the power spectrum and are due to the daily cadence of ground-based observations.} 
The most prominent peaks fall at the rotation frequency and its first harmonic, {as expected} when a few dark spots dominate the RV variations (cf. Sect.~\ref{illustrative_examples} and Figs.~\ref{rv_perturb} and~\ref{conv_shifts}).  This is mainly due to the  convective redshifts in active regions because RV time series computed without their effects have a mean amplitude of only $\sim 18$ m~s$^{-1}$ and an average  power spectrum peaking at the first harmonic with a  power $\sim 2.5$ times lower at the rotation frequency. This happens because all the other perturbations change sign when an active region transits from the approaching to the receding part of the stellar disc, while convective redshifts are always positive. Note that at the orbital frequency of CoRoT-7b and CoRoT-7c, the power is smaller by factors of $\sim 3$ and $\sim 4$ relative to the rotation frequency, respectively. 

{Since the HARPS data are not simultaneous with  CoRoT data, we cannot compare this power spectrum with the periodogram of the HARPS RV time series.}
Nevertheless, it is of interest to apply the technique proposed by \citet[][ Sect.~4]{Quelozetal09} to filter out RV variations due to stellar activity to see to what extent it is capable of  removing the simulated variations associated with stellar surface inhomogeneities. 

\subsubsection{Filtering out activity-related variations from the synthetic radial velocity curve}

To fit activity-induced RV variations, we model the radial velocity $v_{\rm R}(t_{j})$ at a given time $t_{j}$  as a linear combination of sinusoids at the rotation frequency $\Omega = 2 \pi/P_{\rm rot}$, its first and second  harmonics:
\begin{equation}
v_{\rm R} (t_{j}) = a_{0} + \sum_{m=1}^{3} b_{m} \cos (m\Omega t_{j}) +c_{m} \sin (m\Omega t_{j}), 
\end{equation}
where the free coefficients $a_{0}$, $b_{m}$, and $c_{m}$ are determined by minimizing the chi square between the model and the observations. Specifically, we apply the normal equation method   and use LU decomposition and backsubstitution to solve the corresponding set of linear equations for the model coefficients \citep[][ Sect.~15.4]{Pressetal92}. As in  \citet{Quelozetal09}, we adopt $P_{\rm rot}=24.0$ days (although the simulated time series were computed with a model star rotation period of 23.64 days) and apply the three-harmonic fit to subintervals of 20 days. Following \citet{Quelozetal09}, for each point of the time series several best fits are obtained {by scanning the 20-day window through the full data range. These fits are then averaged to compute the best model for that point.} Finally, the best fit time series is subtracted from the initial time series {to obtain the filtered series. A Lomb periodogram is computed for each thus filtered series. Fig.~\ref{per_rv} shows the power spectrum obtained by averaging the periodograms of all the $10^{5}$ residual time series (dashed line). The power due to stellar activity  is  reduced by factors of $\sim 4$ and $\sim 2$ at the orbital frequencies of CoRoT-7b and CoRoT-7c, respectively. }

{To check whether our simulated RV curve of CoRoT-7 contains spurious oscillations at the orbital frequencies of CoRoT-7b or CoRoT-7c, we fit a trigonometric polynomial of the form $B \sin (\omega_{\rm orb} t) + C \cos (\omega_{\rm orb} t)$ to each residual time series and compute the corresponding semiamplitude $A(\omega_{\rm orb})= \sqrt{B^{2} + C^{2}}$, where $\omega_{\rm orb}$ is the orbital frequency of the corresponding planet.} 
{ In the considered $10^{5}$ simulations, $A$ is always lower than $0.85$ m s$^{-1}$ at the frequency of CoRoT-7b and lower than $ 1.7$ m s$^{-1}$ at the frequency of CoRoT-7c. 
The RV oscillations attributed to CoRoT-7b and CoRoT-7c have semiamplitudes $K = 3.5 \pm 0.6$ and 
$K = 4.0 \pm 0.5$ m~s$^{-1}$, respectively. Therefore, the probability that they come from the residual effects of stellar activity is below $10^{-4}$ in both cases. }

{Note that, since the amplitude of the light modulation during the HARPS observing season was greater by a factor of $\sim 1.6$ than during CoRoT observations, the activity-induced  RV oscillations are expected to have been greater during those observations than those simulated from our spot modelling maps.}
Nevertheless, even increasing the amplitudes of the residual oscillations by a factor of 2, the false alarm probability remains below $10^{-4}$. 

\subsubsection{Constraining the facular area}

{While our photometric modelling does not exclude a facular-to-spotted area ratio $Q > 1$, we can constrain it on the basis of the observed RV curve.} 
By computing spot models with $Q$ ranging from 4 to 10, we find that the peak-to-peak amplitude of the RV variations increases from $45.7$ to $120.3$ m~s$^{-1}$ for a light curve relative amplitude of $0.0185$ and spots located within $\pm 20^{\circ}$ from the equator. Given an observed amplitude of $\sim 40$ m~s$^{-1}$ for a photometric relative modulation of $\sim 0.03$, we conclude that the amplitude of the observed RV curve can be reproduced only if $Q < 4$ which {is compatible with our adoption of $Q=1$} corresponding to the minimum of the $\chi^{2}$ of the  composite photometric best fit. {Since the RV amplitude is  dominated by  convective redshifts in faculae and spots, it could be possible to reduce it for $Q \geq 4$ by reducing $\Delta V_{\rm s}$ and $\Delta V_{\rm f}$, or decreasing the spot contrast $c_{\rm s}$ that leads to smaller spots and hence facular areas for a fixed $Q$.  However, the required decrease of $\Delta V_{\rm s}$ and $\Delta V_{\rm f}$ should be at least of $\sim 30$ percent for $Q \sim 6$, i.e., significantly greater than allowed by the scaling relationship proposed by \citet{Gray09}. Similarly, the required decrease of $c_{\rm s}$ should be too large to be compatible with our assumption of active regions with solar-like contrasts. } 

Our model assumes that the active regions are located close to the equator, according to the ME prescription that tries to minimize their area as much as possible. Nevertheless, we also compute a spot model with $Q=6.5$ and the  {constraint} that active regions are located  at high latitudes ($\geq 60^{\circ}$), to study the impact  on the RV  perturbations. {The resulting RV time series is very similar, with differences never exceeding $\pm 6 $ m s$^{-1}$, to the case of low-latitude spots. }  This happens because the reduction of the  convective redshifts due to projection effects is compensated by an increase of the continuum flux perturbation due to the faculae since they are closer to the limb for most of the time. 
{In summary, we conclude that a model with $Q=1$ is appropriate for CoRoT-7, even if its active regions were not constrained to low latitudes.}

\section{Discussion}

{We analyse the white band light curve of the active late-type star CoRoT-7 that hosts a transiting 
telluric planet; a detailed analysis of its RV variations additionally provided evidence of a second non-transiting telluric planet \citep{Legeretal09,Quelozetal09}. Since the major source of RV variations in CoRoT-7 is stellar magnetic activity, it needs to be modelled in order to refine the planetary orbital parameters and confirm the presence of the second planet CoRoT-7c \citep{Quelozetal09}. }
In principle, the RV  oscillation attributed to  CoRoT-7c could be {spurious and due to the effects of}  stellar activity because its  frequency is close to the fifth harmonic of the {stellar} rotational frequency. Unfortunately, the CoRoT observations are not simultaneous to the RV  time series by \citet{Quelozetal09} so our conclusions based on the modelling of CoRoT data do not constrain, rigorously speaking,  magnetic activity properties during the RV observations. {Nevertheless, we can derive illustrative conclusions from our results which can be applied to the analysis of the RV time series. }

Our favoured spot model has a facular-to-spotted area ratio  $Q=1$, as expected for a star with an activity level significantly higher than the Sun. { In our simulated RV time series, most of the power  falls at the rotation frequency and its first  harmonic.} In their analysis of the observed RV time series of CoRoT-7, \citet{Quelozetal09}
 find that most of the power  falls at the rotation frequency and at frequencies corresponding to periods of 9.03 and 10.6 days, the latter corresponding to the first harmonic of the rotation frequency; another prominent peak corresponds to a period of 27.6 days (cf. their Table~2). The period of 27.6 days could be associated with  the rotation of spots at high latitudes given that the relative amplitude of the surface differential rotation of CoRoT-7 is at least of $\sim 6$ percent. {As discussed in Sect.~\ref{spot_model_res}}, a pole-equator angular velocity difference comparable to the Sun, i.e., $\approx 20$ percent, can indeed be present in our star. This is compatible with a rotation period of 23.6 days at the equator and of 27.6 days at high latitudes.  The observed peak corresponding to a period of 9.03 days may therefore be the second harmonic of the high-latitude rotation  frequency.  Recently, Hatzes et al. {(private comm.)} suggested that such a modulation may be associated with a third planet. Our approach cannot be applied to support or disprove this hypothesis because our RV model does not include the effect of surface differential rotation and high-latitude active regions that may be producing the 27.6 day periodicity and its harmonics. Therefore, we have to postpone an investigation of the role of activity-induced effects for the 9.03 day periodicity to a future work.

The presence of a large facular component ($Q \geq 4$) is disfavoured in the case of CoRoT-7 because the amplitude of the RV variation, {in this case dominated by convective redshifts in faculae, would then be too large to be compatible with the RV observations.} 

Finally, we note that in the periodogram of the residuals of our ME composite best fit there is a significant peak corresponding to a period close to the orbital period of CoRoT-7c (cf. Fig.~\ref{res_period}). This peak is  also found  in the periodogram of the time series of the residuals of the best fit obtained with the method of \citet{Quelozetal09}, so it is not an artifact associated with the ME regularization. The corresponding RV  semiamplitude, estimated with the method described in Sect.~\ref{rv_from_model}, is $\sim 0.4$ m~s$^{-1}$, i.e., about one order of magnitude smaller than the RV oscillation produced by the planet, so these fluctuations cannot account for the planetary signature. 
{On the other hand, it cannot be excluded that those  photometric oscillations, with a semiamplitude of only $\sim 6 \times 10^{-5}$ mag, could be induced by the interaction of the planet with the stellar coronal magnetic field or by a perturbation of the stellar dynamo by the planet, as conjectured by, e.g., \citet{Lanza08,Lanza09}. }

\section{Conclusions}
\label{conclusions}

We have applied the spot modelling method introduced by \citet{Lanzaetal09a,Lanzaetal09b} to the {lightcurve of the } planet-hosting star CoRoT-7. Adopting a facular-to-spotted area ratio $Q=1$, we have found evidence of three persistent active longitudes within which individual active regions form and decay with lifetimes ranging from a couple of weeks to a few months. The  active longitudes migrate in the adopted reference frame at different rates which suggests  a surface differential rotation with a relative amplitude of at least $0.058 \pm 0.017$. {The overall spotted area  reaches a maximum of 1.6 percent of the whole surface if the contrasts of dark spots and faculae are adopted as solar-like. }
The residuals of the best fit  indicate a population of  small spots with areas of the order of $5\times 10^{-4}$ of the whole stellar surface that evolve on a typical timescale of $\sim 4-5$ days. Moreover, we find  evidence of an oscillation of the optical flux {with} a semiamplitude $\sim 6 \times 10^{-5}$ mag with the orbital period of the planet CoRoT-7c.

We introduce a  model to simulate the apparent RV variations induced by magnetic activity which includes the effect of surface brightness inhomogeneities as well as a reduction of the macroturbulence and convective blueshifts in  active regions. We synthesize a time series for the RV variations of CoRoT-7 adopting the spot model obtained from CoRoT white band light curve and assuming that active regions have solar-like contrasts and are located within $\pm 20^{\circ}$ from the equator. Unfortunately, HARPS and CoRoT observations are not simultaneous, so any model based on the latter does not constrain, rigorously speaking, the observed RV  variations.  Nevertheless, {an extrapolation of our model matches the observed RV amplitude.} 

We fit the RV  time series with a linear
combination of three sinusoids at the rotation frequency and its first two harmonics, following  \citet[][ Sect.~4]{Quelozetal09}, and confirm that their method is appropriate to filter out most of the activity-induced RV  variations. The residual amplitudes at the orbital periods of the planets CoRoT-7b and CoRoT-7c are  used to estimate false-alarm probabilities that the RV oscillations attributed to the planets are spurious effects induced by stellar activity.
For both planets we find a false-alarm probability lower than $10^{-4}$. 

The possible presence of a third planet with a period of 9.03 days cannot be confirmed by means of our approach because this period could be associated to an harmonic of the signal with a period of 27.6 days which can be produced by spots at high latitude on a differentially rotating star. We shall consider the effects of surface differential rotation in future models of the RV variations to investigate this issue.

\begin{acknowledgements}
The authors are grateful to an anonymous Referee for a careful reading of the manuscript and several suggestions  to improve their work. 
AFL wishes to thank Drs. A.-M.~Lagrange, N.~Meunier, and M.~Desort for interesting discussions. 
Active star research and exoplanetary studies at INAF-Osservatorio Astrofisico di Catania and Dipartimento di Fisica e Astronomia dell'Universit\`a degli Studi di Catania 
 are funded by MIUR ({\it Ministero dell'Istruzione, dell'Universit\`a e della Ricerca}) and by {\it Regione Siciliana}, whose financial support is gratefully
acknowledged. 
This research has made use of the CoRoT Public Data Archive operated at IAS, Orsay, France, and of the ADS-CDS databases, operated at the CDS, Strasbourg, France.
\end{acknowledgements}

\end{document}